\documentclass[letterpaper, 10 pt, conference]{ieeeconf}  

\IEEEoverridecommandlockouts                              

\usepackage{amsmath} 
\usepackage{amssymb}  
\usepackage{booktabs}

\title{\LARGE \bf
	Switched Lyapunov Function based Controller Synthesis for Networked Control Systems: A Computationally Inexpensive Approach
}

\author{Katarina Stanojevic$^{1}$, Martin Steinberger$^{1}$ and Martin Horn$^{1,2}$
\thanks{$^{1}$ Katarina Stanojevic, Martin Steinberger and Martin Horn are with the Institute of Automation and Control, Graz University of Technology, Graz, Austria. {\tt katarina.stanojevic@tugraz.at}}%
\thanks{$^{2}$ Martin Horn is with the Christian Doppler Laboratory for Model Based
	Control of Complex Test Bed Systems, Institute of Automation and Control, Graz University of Technology, Graz, Austria.\newline
	The financial support by the Christian Doppler Research Association, the
	Austrian Federal Ministry for Digital and Economic Affairs and the National
	Foundation for Research, Technology and Development is gratefully
	acknowledged.}
}

\usepackage{bm}
\usepackage{pdflscape}
\usepackage{mathtools}
\usepackage{adjustbox}

\usepackage{tikz}
\usetikzlibrary{shapes,arrows}
\usepackage{circuitikz}
\usepackage{pgfplots}
\usepackage{makecell}
\definecolor{tugred}{RGB}{247,1,70}
\definecolor{tugblue}{RGB}{17,71,132}
\definecolor{tugorange}{rgb}{0.97,0.59,0.27}
\definecolor{tuggreen}{rgb}{0.61,0.73,0.35}
\definecolor{tugmagenta}{rgb}{0.50,0.39,0.64}
\definecolor{mygreen}{RGB}{28,172,0}

\usepackage{multirow}
\usepackage{pgfplots} 
\begin{document}

\maketitle
\thispagestyle{empty}
\pagestyle{empty}

\begin{abstract}
This paper presents a Lyapunov function based control strategy for networked control systems (NCS) affected by variable time delays and data loss. A special focus is put on the reduction of the computational complexity. A specific buffering mechanism is defined first, such that it adds an additional delay up to one sampling period. The resulting buffered NCS can then be formulated as a switched system which leads to the significant simplification of the NCS model and subsequent controller synthesis. The novel approach does not only circumvent the need for any over-approximation technique, since the switched NCS model can be used for stability analysis directly, but also reduces the infinite set of allowable values of the dynamic matrix to a small finite set. The proposed strategy leads hereby to a strongly decreased number of optimization variables and linear matrix inequalities (LMIs) which allows greater flexibility with respect to additional degrees of freedom affecting the transient behavior. The performance and computational efficiency of the control strategy are demonstrated by means of simulation example.
\end{abstract}

\tikzstyle{block4} = [draw,thick, rectangle, 
minimum height=4em, minimum width=3em, rounded corners]
\tikzstyle{block5} = [draw,thick, rectangle, 
minimum height=3em, minimum width=8em, rounded corners]
\tikzstyle{block} = [draw,thick, thick, rectangle, 
minimum height=3em, minimum width=4em,rounded corners]
\tikzstyle{block2} = [draw,thick, rectangle, 
minimum height=5.0em, minimum width=2em, rounded corners]
\tikzstyle{block3} = [draw,thick, rectangle, 
minimum height=3em, minimum width=3em, rounded corners]
\tikzstyle{block33} = [draw,thick, rectangle, 
minimum height=2.5em, minimum width=2.5em, rounded corners]
\tikzstyle{network} = [draw,thick, color=red, rectangle, 
minimum height=3em, minimum width=3em, rounded corners]
\tikzstyle{sum} = [draw,thick, circle, node distance=1cm]
\tikzstyle{input} = [coordinate]
\tikzstyle{output} = [coordinate]
\tikzstyle{pinstyle} = [pin edge={to-,thin,black}]

\section{Introduction}
The great potential and necessity for wireless networked control systems (NCS) which are characterized by controlling the plant over a (shared) communication channel together with significant developments of communication technologies have moved these systems to the center of attention within many engineering industries. This has been motivated by increasing complexity of modern control systems leading to the rapidly growing interest in characteristics the NCSs possess, such as great flexibility and adaptability regarding the system topology. Nevertheless, the main challenges which arise when controlling the system over a network are communication constraints such as limited transmission speed and unreliability of the communication channel. The resulting random time delays and data loss have a great impact on the control design, since they don't in general allow a direct application of the conventional control laws. This leads to the need for new strategies which take these into account and ensure the stability and desired performance of the NCS despite the network uncertainties. An overview of the extensive literature considering the modeling of the network imperfections, the stability properties and control of NCS can be found in, e.g., \cite{Zhang2016}, \cite{Parks2018} and references therein.

The Lyapunov stability theory and analysis of dynamical systems in terms of Linear Matrix Inequalities (LMIs) has been essential in the control systems community, since it was proposed in literature \cite{lmibook}. The possibility to express a control problem as a standard convex optimization problem involving LMIs which can efficiently be solved numerically has been recognized in the field of NCS, as well. A very powerful stability characterization based on parameter-dependent Lyapunov functions is proposed in \cite{PosthumusCloosterman2008}, \cite{Cloosterman2010}, which leads to a control law ensuring the stability of NCS including time-varying delays, packet dropouts and variations in the sampling interval. However, the stability analysis and controller synthesis for time-variant NCS models combining all relevant network uncertainties results in a very high complexity of this method. First, a suitable polytopic model must be derived using over-approximation techniques such as the Jordan Form. Second, the number of LMIs which needs to be solved to compute a control law is very high and increases exponentially when, e.g., the maximum allowable delay is increased.

To circumvent the high computational complexity of the aforementioned approach, a buffering mechanism has been proposed in literature, see, e.g., \cite{Ludwiger2018}, \cite{Ludwiger2019}, \cite{Stanojevic2022} which adds additional delay so, that a constant (worst-case) delay is ensured. The great advantage of this approach is that the resulting NCS model becomes time-invariant which significantly simplifies the control design.  Nevertheless, even though the delay introduced by the network might be short for some data packets, they are not forwarded to the plant until the maximal delay is reached, therefore leading often to unnecessarily long delays, which are in general undesirable. 

In this paper, we propose an LMI-based control strategy suitable for NCS affected by variable time delays and packet dropouts motivated by \cite{Cloosterman2010} but with a special focus on the reduction of the complexity of the method. The simplification of the NCS model is achieved by a specific buffering mechanism \cite{Stanojevic2022ccta}, which significantly reduces additional buffer delays in comparison to the worst-case scenario from \cite{Ludwiger2018}. The resulting buffered NCS can be therefore formulated as a switched system, for which an extensive literature related to LMI-based control design is available, see, e.g., \cite{deOliveira1999}, \cite{Daafouz2001}, \cite{Daafouz2002}. This represents a crucial step toward the computational simplicity, since it circumvents the need for over-approximation techniques and strongly reduces the number of optimization variables and resulting LMIs. Furthermore, the significant reduction of the algorithms' complexity allows the introduction of the additional degrees of freedom, which can be used to influence the transient behavior of the system, increasing the practical usability of the strategy. The proposed control laws are compared to the control law from \cite{Cloosterman2010} in simulations on the basis of TrueTime \cite{Cervin2003} with respect to their performance and computational complexity. 

\section{Problem Formulation}
\begin{figure}
	\centering
	\medskip
	\resizebox{0.49\textwidth}{!}{%
		\begin{tikzpicture}[auto, node distance=2cm,>=latex']
			\node [input, name=input] {};
			\node [block2, right of=input, node distance=1cm,inner xsep=-0.13cm] (buffer) {{$\begin{array}{cc} \large \text{Buffer}\\ \large \tau_k^b \end{array}$ }};
			\node [block2, right of=buffer, node distance=1.4cm,inner xsep=0.03cm]   (zoh) {\large ZOH};
			\draw [thick,->] (buffer) -- (zoh);
			\draw[tugred,very thick, rounded corners,label=90:\large actuator] ([yshift=9.5mm,xshift=5mm]buffer.west)-|([xshift=0.5mm]zoh.east)|-([yshift=-9.5mm,xshift=-0.5mm]buffer.west) |- ([yshift=9.5mm,xshift=5mm]buffer.west);
			\node[text=tugred] at (1.7,11.5mm) {\large actuator node};
			
			\node [block2,inner xsep=-0.1cm,minimum height=5.2em] at (5,0)  (plant) {$\begin{array}{cc} \large \text{Cont.-time}\\ \large \text{plant} \end{array}$};
			\begin{scope}[transform canvas={yshift=0.7cm}]
				\draw [->] (zoh) -- node {\large$u_1^*(t)$} (plant);
			\end{scope} 
			\begin{scope}[transform canvas={yshift=-0.7cm}]
				\draw [->] (zoh) -- node {\large$u_m^*(t)$} (plant);
			\end{scope}
			\node[text = black] at (3.2, 0.35) {{ $\bm{\vdots}$}};
			
			\node [block,label=-90:\large sensor node] at (8.5,0)  (sensor) {}; 
			\draw [thick,->] (plant) -- node { \large $\bm{x}(t)$} (sensor) ;
			
			\draw[thick] (7.35,0) -- (8.05+0.1,0);
			\draw[thick] (8.1+0.1,0) -- (8.5+0.1,0.25);
			\draw[thick, fill=white] (8.1+0.1,0)  circle[radius=1.4pt];
			\draw[thick, fill=white] (8.6+0.1,0)  circle[radius=1.4pt];
			\draw[thick] (8.75,0) -- (10,0);
			
			\node [block3] at (5,-4.2) (controller) {$\begin{array}{cc}\large \text{Discrete time}\\ \large \text{controller} \end{array}$};
			\node [block3] at (3.0,-4.2) (tau3) {\large $\tau_k^{c}$};
			
			\node [output, right of=sensor, node distance=1.5cm] (output) {}; 
			\node [block33]  at (10,-1.6) (tau1) {\large$\tau_k^{sc}$};
			\draw [->,thick] (output) -- (tau1);
			\draw [thick] (sensor) -- node {\large $\bm{x}_k $} (output) ;
			\node [block33,minimum height=2.8em, label = 180: $m_k^{sc}$]  at (10,-2.9) (m1) {};
			\draw[thick] (10,-2.4) -- (10,-2.6);
			\draw[thick] (10,-3.2) -- (10,-3.5);
			\draw[thick, fill=white] (10,-2.65)  circle[radius=1.4pt];
			
			\draw[thick] (9.7,-2.8) -- (10,-3.15);
			\draw[thick, fill=white] (10,-3.15)  circle[radius=1.4pt];
			
			\draw [->,thick] (tau1) -- (m1);
			\draw [->,thick] (m1) |- (controller);
			\draw [->,thick] (input) -- (buffer);
			\node [block33]  at (0,-1.6) (tau2) {\large $\tau_k^{ca}$};
			\draw [thick] (tau2) |- (input);

			\node [block33,minimum height=2.8em, label = 0: $m_k^{ca}$]  at (0,-2.9) (m2) {};
			\draw[thick] (0,-2.4) -- (0,-2.6);
			\draw[thick] (0,-3.2) -- (0,-3.5);
			\draw[thick, fill=white] (0,-2.65)  circle[radius=1.4pt];
			
			\draw[thick] (-0.3,-2.8) -- (-0,-3.15);
			\draw[thick, fill=white] (0,-3.15)  circle[radius=1.4pt];
			
			\draw [<-,thick] (tau2) -- (m2);
			\draw [->,thick] (tau3) -| (m2);
			
			\draw[tugblue,very thick, rounded corners,opacity=.8,dashed] ([yshift=5mm,xshift=3mm]tau2.west)-|([xshift=3mm]tau1.east)|-([yshift=-5.5mm,xshift=-3mm]m2.west) |- ([yshift=5mm,xshift=3mm]tau2.west);
			\node[text=tugblue] at (5,-2.2) {{\Large Network}};
			
			\draw [->,thick] (10.5-1.5,1.2) -- (10.5-1.5,0.5);
			\node[text=black] at (10.8-1.5,1.2) {\large $T_d$};
		\end{tikzpicture}
	}%
	\caption{Considered buffered networked control system}
	\label{ncs}
\end{figure}
In this work we consider systems which can be described by a continuous-time multi-input linear time-invariant model
\begin{equation}
	\dot{\bm{x}}(t) = \bm{A}_c\bm{x}(t)+\bm{B}_c \bm{u}^*(t)
	\label{cont.plant}
\end{equation}
with state vector $\bm{x}(t)\in \mathbb{R}^n$ and input $\bm{u^*}(t)=\begin{bmatrix}
u_1^* & u_2^* &\dots & u_m^*
\end{bmatrix}^T \in \mathbb{R}^m $. The system to be controlled including sensor and actuator nodes at the plant side is connected over a wireless communication network to the controller side, where a discrete-time controller providing actuating signals for all $m$ input channels is implemented, see Fig. \ref{ncs}. In the sensor node, the discrete time signal $\bm{x}_k$ is generated with a constant sampling period $T_d$. The $k^{th}$-data packet containing the sampled states $\bm{x}_k$ and a timestamp $t = kT_d$ ($k \in \mathbb{N}_0$) is sent over a network to the controller. Based on the received data the discrete-time control signal $\bm{u}_k$ is computed and forwarded over a network back to the plant side, where a zero-order hold block (ZOH) is used to generate the actuation signal $\bm{u}^* (t)$. Additionally, a message rejection mechanism is implemented in the actuator node, which ensures that no older control data is applied to the system. The possibly negligible variable time needed for the execution of the control law is included by the variable $\tau_k^c$.  Furthermore, the aforementioned network imperfections considered in this work are modeled by variable network-induced sensor-to-controller delays $\tau_k^{sc}$ and controller-to-actuator delays $\tau_k^{ca}$. In addition, the possibility of lossy networks is incorporated by variables $m_k^{sc}$ and $m_k^{ca}$ which denote whether the $k^{th}$ data packet is lost ($m_k = 1$) or received ($m_k = 0$). For the networked system under consideration, the following assumptions hold:

\textit{Assumption 1:} 	System \eqref{cont.plant} is controllable. The controllability is not lost due to sampling, i.e. the sampling time $T_d$ is non-pathological in the sense of \cite{Kalman2011}.

\textit{Assumption 2:} The individual delays $\tau_k^{sc}$, $\tau_k^{ca}$ and $\tau_k^c$ are assumed to be upper bounded (if no data is lost) by $\bar{\tau}^{sc}, \bar{\tau}^{ac}$ and $\bar{\tau}^c$, respectively. The upper bounds can be larger than $T_d$ (large delay case).  A specific distribution of the delays is not considered.

\textit{Assumption 3:}  Data loss in the network is modeled by defining the upper bound $\bar{p}$ for the number of subsequent packet dropouts, i.e. $\sum_{i = k-\bar{p}}^{k} m_i\leq \bar{p}$.  A specific distribution of the packet dropouts is not considered.


The discrete-time controller provides actuating signals $u_{k,i}$ for all $i = 1, 2, \dots, m$ input channels. For the stability analysis and the control design it is necessary to express the control signal $\bm{u^*}(t)$ acting on the plant \eqref{cont.plant} for $kT_d\leq t <(k+1)T_d$ in form of computed discrete-time control inputs $\bm{u}_k$ \cite{Cloosterman2010}. 

The delay $\tau_k$ which denotes the complete time from the moment of sampling of $\bm{x}_k$ at $t = kT_d$ at the sensor node up to the moment when the control signal computed based on $\bm{x}_k$ is available at the actuator depends on the individual delays present in the network, the network structure and the proposed strategy. In this work we first consider static controllers. Therefore, the delay introduced by the network for the $k^{th}$ data packet is given as $	\tau_k = \tau_k^{sc}+\tau_k^c+\tau_k^{ca}$, 
see \cite{Cloosterman2010}, which is bounded due to Assumption 2, i.e. $\exists \bar{d} \in \mathbb{N}^+$ so that ${\tau}_k \leq \bar{d} T_d$ $\forall k$  holds. An additional parameter combining the considered imperfections of the network can, therefore, be defined as
\begin{equation}
	\bar{\delta} = \bar{d}+\bar{p},
	\label{delta}
\end{equation}
which allows expressing the continuous-time actuation signal as $\bm{u}^*(t) = \bm{u}_{k+j-\bar{\delta}} \quad \text{for} \quad kT_d+t_k^j\leq t <kT_d +t_k^{j+1} $ ($j = 0,1,\dots, \bar{\delta} $), see \cite{Cloosterman2010}. The parameters $t_k^j$ denote time instants at which the corresponding $\bm{u}_{k+j-\bar{\delta}}$ become available at the actuator,  as defined in \cite{PosthumusCloosterman2008}, \cite{Cloosterman2010}. The discretization of \eqref{cont.plant} with a constant sampling time $T_d$ leads to the discrete time model
\begin{equation}
	\bm{x}_{k+1} = \bm{A}_d\bm{x}_k +\sum_{j = 0}^{\bar{\delta}} \bm{B}_j(t_k^j,t_k^{j+1}) \bm{u}_{k+j-\bar{\delta}}
	\label{disc.plant}
\end{equation}
with $\bm{B}_j (t_k^j,t_k^{j+1}) = \int_{t_k^j}^{t_k^{j+1}} e^{\bm{A}_c\left(T_d-s\right)}\bm{B}_c ds
$
and $\bm{A}_d = e^{\bm{A}_cT_d}$. From the complexity of the resulting time-variant discrete-time model \eqref{disc.plant}, which arises as a consequence of purely random network effects, it is clear that the conventional control approaches cannot be directly applied. The aim of this paper is to define a control strategy suitable for the NCS affected by random network imperfections with a special focus on the practicability of the method and reduction of the computational complexity.

\section{Proposed Approach}
In the following, a switched Lyapunov function based control strategy is presented.
Firstly, in order to simplify the discrete time-model \eqref{disc.plant} and avoid the necessity for the convex over-approximation and a large number of linear matrix inequalities (LMIs) in the subsequent controller design, a buffering mechanism is defined. Based on this simplification, an LMI-based control law is proposed, with the focus on the practical usability in terms of computational simplicity and flexibility of the approach.
\subsection{Buffering Mechanism}
The proposed buffering mechanism is implemented as a part of the actuator node at the receiving end of the feedback channel at the plant side, see Fig. \ref{ncs}. Its effect is defined such, that, if a data packet arrives at the plant side, the buffer adds an additional delay $\tau_k^b$ before forwarding the control signal to the ZOH, i.e. the overall delay is reformulated to $\tau_k = \tau_k^{sc}+\tau_k^{c}+\tau_k^{ca}+\tau_k^{b}$. The delay $\tau_k^b$ is calculated based on the timestamp attached by the sensor, so that $
	\tau_k = \left\lceil \frac{\tau_k^{sc}+\tau_k^{c}+\tau_k^{ca}}{T_d}\right\rceil \cdot T_d = q_k T_d$
holds. The buffer delay is therefore given as $\tau_k^b = q_k T_d - (\tau_k^{sc}+\tau_k^{c}+\tau_k^{ca}) < T_d$. This ensures that the control signal is updated at $t = kT_d$ and that there is only one signal acting on the plant during one sampling period. In the case when more than one data packet arrive during one sampling period, the most recently received control signal is applied, whereas in the case when no new signal is received, the previously applied $\bm{u}_k$ is kept active. Even though it is still unknown when a specific control signal will be applied to the system, the minimal time duration for which an applied control input is active is defined explicitly. The NCS can be therefore described by a simplified, discrete-time model
\begin{equation}
	\bm{x}_{k+1} = \bm{A}_d\bm{x}_k + \bm{B}_d \bm{u}_{k-q_k}
	\label{disc.plant_buffer}
\end{equation}
with 
	$\bm{B}_d = \int_{0}^{T_d} e^{\bm{A}_c\left(T_d-s\right)}\bm{B}_c ds = \begin{bmatrix}
		\bm{b}_1 & \bm{b}_2 & \dots & \bm{b}_m
	\end{bmatrix}$
and $q_k \in \{1,2, \dots,\bar{\delta}\}$ being the unknown input delay which depends on the network imperfections, see Assumptions 2 and 3.
A model suited for stability analysis is obtained by defining the following lifted-state vector 
\begin{equation}
	\begin{aligned}
		\bm{\xi}_k &=
		\left[\begin{matrix}
			\bm{x}_k^T & u_{1,k-1} & u_{1,k-2} & \dots & u_{1,k-\delta} & \dots \\
		\end{matrix}\right.\\
		&\qquad\qquad
		\left.\begin{matrix}
			& & u_{m,k-1} & u_{m,k-2} & \dots &u_{m,k-\delta}
		\end{matrix}\right]^T
	\end{aligned}
	\label{liftedvector}
\end{equation}
which results in the lifted model 
\begin{equation}
	\bm{\xi}_{k+1} = \bm{\hat{A}}(\bm{\alpha}_k)\bm{\xi}_k+\bm{\hat{B}}\bm{u}_k.
	\label{lifted}
\end{equation}
The indicator function
\begin{equation}
\bm{\alpha}_k = \begin{bmatrix}
\alpha_{1,k} & \alpha_{2,k} & \dots & \alpha_{\bar{\delta},k}
\end{bmatrix}^T
\label{alpha}
\end{equation}
is thereby defined as $\alpha_{i,k} = 1$ if $\bm{u}_{k-i}$ is active for $kT_d \leq t <(k+1)T_d$ and $\alpha_{i,k} = 0$ otherwise,
with $\sum_{i=1}^{\bar{\delta}} \alpha_{i,k} = 1$. The matrices $\bm{\hat{A}}(\bm{\alpha}_k)$ and $\bm{\hat{B}}$ are given as
\begingroup
\allowdisplaybreaks
	\begin{align}
		&\hat{\bm{A}}(\bm{\alpha}_k) = \begingroup 
		\setlength\arraycolsep{2pt} \begin{bmatrix*}[c]
			\bm{A_d}&\alpha_{1,k} \bm{B}_d \, \dots \, \alpha_{\bar{\delta}-1,k} \bm{B}_d & \alpha_{\bar{\delta},k} \bm{B}_d\\
			\bm{0}_{m\times n} & \hspace*{-2mm}\bm{0}_{m} \,\,\dots \,\,\,\, \bm{0}_{m} &\bm{0}_{m} \\
			\bm{0}_{(\bar{\delta}-1)m\times n} & {\bm{I}_{(\bar{\delta}-1)m}} & \bm{0}_{(\bar{\delta}-1)m\times m} \\
		\end{bmatrix*}\hspace*{-1.5mm},\endgroup\label{Ahat} \\
		&\hat{\bm{B}} = \begin{bmatrix*}[l]
			\bm{0}_{m\times n } & \bm{I}_{m \times m} & \bm{0}_{m\times (\bar{\delta}-1)m }
		\end{bmatrix*}^T = \begin{bmatrix*}[l]
			\hat{\bm{b}}_1&\hat{\bm{b}}_2\,\,\dots\,\,\hat{\bm{b}}_m
		\end{bmatrix*}.\nonumber
	\end{align}
Due to the fact that there is only one signal acting between two sampling instants, the lifted model \eqref{lifted} can easily be written as a switched system
\begin{equation}
	\bm{\xi}_{k+1} = \sum_{i = 1}^{\bar{\delta}}\alpha_{i,k}\bm{\hat{A}}_i\bm{\xi}_k+\bm{\hat{B}}\bm{u}_k
	\label{switched}
\end{equation}
with a switching rule contained in \eqref{alpha}, which is not known \textit{a priori}, since it depends on the unknown network delays. The matrices {\small$ \bm{\hat{A}}_i$} in \eqref{switched} can be determined by evaluating \eqref{Ahat} for  $\bm{\alpha_}{k} = \bm{e}_i$ with $\bm{e}_i$ being the vector whose components are all zero, except the $i^{th}$ element which equals $1$. Therefore, the usually complex convex over-approximation of the NCS model used to obtain the polytopic model which is suitable for stability analysis and control synthesis \cite{PosthumusCloosterman2008}, \cite{Cloosterman2010} for \eqref{disc.plant} is no longer necessary. The infinite set of allowable values of the dynamic matrix is hereby reduced to a finite set containing only $\bar{\delta}$ elements.

\subsection{LMI based Controller for Switched NCS}
In this section we propose a static state feedback control law of the form 
\begin{equation}
\bm{u}_k = -{\bm{K}}_x\bm{x}_k = -\hat{\bm{K}}\bm{\xi}_k = -\begin{bmatrix}
\bm{K}_{x} & \bm{0}
\end{bmatrix}\bm{\xi}_k,
\label{controller}
\end{equation}
 which ensures asymptotic stability of the switched hybrid system \eqref{switched} leading to the closed loop system
 \begin{equation}
 	\bm{\xi}_{k+1} = \left( \sum_{i = 1}^{\bar{\delta}}\alpha_{i,k}\bm{\hat{A}}_i - \bm{\hat{B}\hat{\bm{K}}}\right) \bm{\xi}_k.
 	\label{closed_lifted}
 \end{equation} The motivation to define the controller \eqref{controller} as a linear combination of system states only, is based on the need to avoid more restrictive assumptions and possible deadlocks which could occur when using a dynamic controller, see \cite{Cloosterman2010}.

\textit{Theorem 1:}
Consider the buffered NCS \eqref{disc.plant_buffer} with a static state feedback controller \eqref{controller} and let Assumption 1-3 hold. If there exist symmetric positive definite matrices $\bm{Y}_i \in \mathbb{R}^{(n+\bar{\delta} m) \times (n+\bar{\delta} m)}$, a matrix $\bm{Z} \in \mathbb{R}^{m \times n}$, matrices $\bm{X}_i = \begin{bmatrix}
	\bm{X}_{1} & \bm{0}_{n \times \bar{\delta} m} \\ \bm{X}_{2,i} & \bm{X}_{3,i}
\end{bmatrix}$ 
with $\bm{X}_1 \in \mathbb{R}^{n\times n}$, $\bm{X}_{2,i} \in \mathbb{R}^{\bar{\delta} m\times n}$, $\bm{X}_{3,i} \in \mathbb{R}^{\bar{\delta} m\times \delta m}$ for $i \in \{0,1 \dots,\bar{\delta}\}$ and a scalar $0\leq\gamma<1 $ such that the linear matrix inequality 
\begin{equation}
\begin{bmatrix}
	\bm{X}_i+\bm{X}_i^T-\bm{Y}_i & \bm{X}_i^T\bm{\hat{A}}_i^T-\bm{\hat{Z}}^T\bm{\hat{B}^T} \\
\bm{\hat{A}}_i \bm{X}_i - \bm{\hat{B}} \bm{\hat{Z}} & (1-\gamma) \bm{Y}_j
\end{bmatrix}> \bm{0}
\label{ineq}
\end{equation}
with $\bm{\hat{Z}} = \begin{bmatrix}
	\bm{Z} & \bm{0}_{m\times \bar{\delta}} 
\end{bmatrix}$ is satisfied for $\forall i,j \in \{1,2,\dots,\bar{\delta}\}$, then the buffered NCS is asymptotically stable. Moreover, the state feedback gain matrix is determined by $\bm{K}_x = 	\bm{Z}  \bm{X}_1^{-1}$
and the corresponding Lyapunov function is given as $V(\bm{\xi}_k,\bm{\alpha}_k) = \bm{\xi}_k^T \bm{\hat{P}}(\bm{\alpha}_k) \bm{\xi}_k $
with the parameter dependent Lyapunov matrix $
\bm{\hat{P}}(\bm{\alpha}_k) = \sum_{i = 1}^{\bar{\delta}}\alpha_{i,k}\bm{P}_i $
for $\bm{P}_i = \bm{Y}_i^{-1}$, which is positive definite for all values of the indicator function $\bm{\alpha}_k$ \eqref{alpha}.

\textit{Proof:} The relation between the matrix $\bm{\hat{Z}}$ from \eqref{ineq} and the control gain is given by $\bm{\hat{Z}} = \bm{\hat{K}} \bm{X}_i$
leading to the LMIs
\begin{equation}
\begin{bmatrix}
	\bm{X}_i+\bm{X}_i^T-\bm{Y}_i & \left( \left( \bm{\hat{A}}_i - \bm{\hat{B}} \bm{{\hat{K}}} \right) \bm{X}_i \right)^T \\
	\left( \bm{\hat{A}}_i - \bm{\hat{B}} \bm{{\hat{K}}} \right) \bm{X}_i & (1-\gamma) \bm{Y}_j
\end{bmatrix}> \bm{0},
\label{LMIimplement}
\end{equation}
where the closed loop dynamic matrices for individual modes of the switched system are contained. For $\bm{Y}_i^{-1} = \bm{P}_i$ this is equivalent to 
\begin{equation}
	\begin{bmatrix}
		\bm{P}_i & \left( \bm{\hat{A}}_i - \bm{\hat{B}} \bm{{\hat{K}}} \right)^T \bm{P}_j\\
		\bm{P}_j\left( \bm{\hat{A}}_i - \bm{\hat{B}} \bm{{\hat{K}}} \right)  & (1-\gamma) \bm{P}_j
	\end{bmatrix} = \bm{Q}_{ij}> \bm{0},
\label{Qij}
\end{equation}
see \cite{Daafouz2001}, \cite{Daafouz2002}. From the Schur complement of $\sum_{i = 1}^{\bar{\delta}} \alpha_{i,k}\sum_{j = 1}^{\bar{\delta}} \alpha_{j,k} \bm{Q}_{ij} $, the inequality 
\begin{equation}
\bm{\hat{P}} - \left( \bm{\hat{A}}(\bm{\alpha}_k) - \bm{\hat{B}} \bm{{\hat{K}}} \right)^T \bm{\hat{P}^+} \left( \bm{\hat{A}}(\bm{\alpha}_k)-\bm{\hat{B}} \bm{{\hat{K}}} \right)  > \gamma \bm{\hat{P}}
\label{delgamma}
\end{equation}
can be derived with 
	$\bm{\hat{P}^+} = \bm{\hat{P}}(\bm{\alpha}_{k+1}) = \sum_{j = 1}^{\bar{\delta}}\alpha_{j,k}\bm{P}_j$
and $0 \leq\gamma < 1$,
see \cite{Daafouz2001}. The equilibrium of \eqref{closed_lifted} is globally uniformly asymptotically stable if 
\begin{equation} \label{delV}
	\begin{split}
		-&{\Delta}V(\bm{\xi}_k,\bm{\alpha}_k) = V(\bm{\xi}_k,\bm{\alpha}_k)- V(\bm{\xi}_{k+1},\bm{\alpha}_{k+1}) = \\
		=&\bm{\hat{P}} - \left( \bm{\hat{A}}(\bm{\alpha}_k) - \bm{\hat{B}} \bm{{\hat{K}}} \right)^T \bm{\hat{P}^+} \left( \bm{\hat{A}}(\bm{\alpha}_k)\bm{\hat{B}} \bm{{\hat{K}}} \right)  > \bm{0} 
	\end{split}
\end{equation}
holds. Due to the fact that $\hat{\bm{P}}$ is positive definite, this condition is ensured when \eqref{delgamma} is fulfilled. \hfill $\blacksquare$

The impact of the scalar value $\gamma$ becomes clear from \eqref{delgamma} and \eqref{delV}, since increasing its value results in the larger value of $\left| {\Delta}V(\bm{\xi}_k,\bm{\alpha}_k) \right|$ and therefore leading to the larger lower bound for the transient decay rate of the system states. Hence, the parameter $\gamma$ represents an additional variable which can be used to influence the transient performance and resulting settling time, see \cite{PosthumusCloosterman2008}.

The total number of necessary LMI conditions in the form \eqref{LMIimplement} which should be solved is $\bar{\delta}^2$. This represent a significant result in reduction of the computational complexity in comparison to the approach presented in \cite{PosthumusCloosterman2008} and \cite{Cloosterman2010}, where the number of LMI conditions to be solved is $2^{2\bar{\delta} \nu}$ with $\nu \leq n$ depending on the geometric multiplicity of the eigenvalues, see \cite{PosthumusCloosterman2008}. 
\section{Extensions of the Control Design}
This section addresses extensions of the previously proposed static feedback control law which can be applied to enhance the performance of the complete NCS.
\subsection{Switched State Feedback Control Law}
\textit{Theorem 2:}
Consider the buffered NCS \eqref{disc.plant_buffer} with a switched state feedback controller
\begin{equation}
	\bm{u}_k = -\bm{K}_{x,i} \bm{x}_k = -\hat{\bm{K}}_i\bm{\xi}_k = -\begin{bmatrix}
		\bm{K}_{x,i} & \bm{0}
	\end{bmatrix}\bm{\xi}_k
	\label{controllerI}
\end{equation}
and consider the scalar $\gamma$ and the matrices as in the Theorem 1 with the difference that the constant matrices $\bm{X}_1$ and $\bm{Z}$ are replaced by matrices $\bm{X}_{1,i}$ and $\bm{Z}_i$ of the same size ($\forall i = 1,2,\dots,\bar{\delta}$), respectively. If the resulting LMIs of the form \eqref{LMIimplement} are satisfied $\forall i,j \in \{1,2,\dots,\bar{\delta}\}$, then the buffered NCS is asymptotically stable. 
The switched control law \eqref{controllerI} is obtained by $\bm{\bm{K}_{x,i}} = \bm{Z}_i \bm{X}_{1,i}^{-1}$ leading to $\bar{\delta}$ values of the control signal $\bm{u}_k$ computed based on single data packet containing $\bm{x}_k$. 

\textit{Proof:} The proof follows as a direct consequence of the Theorem 1.  \hfill $\blacksquare$

\noindent Please note that in this case, an additional mechanism implemented in the buffer is necessary. Since it is not known at the moment of the computation of the controller which  $\bm{\hat{A}}_i$ mode is active, the data packet containing all the $\bar{\delta}$ values for $\bm{u}_k$ is sent to the actuator node. Based on the timestamp attached by the sensor, the total delay can be calculated which allows the buffer to determine which control law \eqref{controllerI} should be applied.

\subsection{Extended State Feedback Control Law}
\textit{Theorem 3:}
Consider the buffered NCS \eqref{disc.plant_buffer} with an extended state feedback controller
\begin{equation}
	\bm{u}_k = -\hat{\bm{K}}\bm{\xi}_k = -\begin{bmatrix}
		\bm{K}_{x} & \bm{K}_u
	\end{bmatrix}\bm{\xi}_k
	\label{controllerII}
\end{equation}
which includes not only the state variables but also part of the control input history. Furthermore, consider the scalar $\gamma$ and the matrices as in the Theorem 1 with the difference that the matrices $\bm{X}_i$ and $\bm{\hat{Z}}$ are replaced by one constant matrix $\bm{X}$ and a full matrix $\bm{\hat{Z}} = \begin{bmatrix}
	\bm{Z}_x & \bm{Z}_u
\end{bmatrix}$ ($\forall i = 1,2,\dots,\bar{\delta}$), respectively. If the resulting LMIs of the form \eqref{LMIimplement} are satisfied $\forall i,j \in \{1,2,\dots,\bar{\delta}\}$, then the buffered NCS is asymptotically stable for the extended control law \eqref{controllerII} obtained by $\bm{K} = \bm{\hat{Z}}\bm{X}^{-1}$.

\textit{Proof:} The proof follows as a direct consequence of the Theorem 1. \hfill $\blacksquare$

Please note that if an extended control law is used, an  additional restriction that there are no dropouts in the connection path from the plant to the controller must hold. In addition, the controller must wait until all the previous control signals are available before computing $\bm{u}_k$, which could add an additional delay in the system and hence leads to the reformulation of the sensor-to controller delay to $\tau_k^{sc} = \max \{kT_d+\tau_k^{sc}, jT_d + \tau_j^{sc}\}, \forall j < k $. However, the maximal round trip time and therefore the value of $\bar{\delta}$ \eqref{delta} remain unchanged, see Assumption 2, allowing that all the considerations  directly follow from Theorem 1.
\subsection{Additional Influence on the Transient Performance}
The advantage of the previously proposed control laws regarding the significantly reduced computational effort allows further extensions of the control law proposed in Theorem 1, 2 and 3 by introducing additional degrees of freedom. This is done by replacing one scalar $\gamma$ by $\gamma_i$  for $i = 1,2,\dots \bar{\delta}$ which have impact on the transient performance of the NCS. 
 \textit{Theorem 4:} Consider the buffered NCS \eqref{disc.plant_buffer} with a static state feedback controller \eqref{controller} and the matrices $\bm{Y}_i$, $\bm{X}_i$ and $\bm{Z}$ as in the Theorem 1. Furthermore, consider scalars $0 \leq \gamma_i < 1$ ($i = 1,2,\dots, \bar{\delta}$) such that the linear matrix inequality 
 \begin{equation}
 	\begin{bmatrix}
 		\bm{X}_i+\bm{X}_i^T-\bm{Y}_i & \bm{X}_i^T\bm{\hat{A}}_i^T-\bm{\hat{Z}} ^T\bm{\hat{B}^T} \\
 		\bm{\hat{A}}_i \bm{X}_i - \bm{\hat{B}} \bm{\hat{Z}} & (1-\gamma_i) \bm{Y}_j
 	\end{bmatrix}> \bm{0}
 	\label{ineq2}
 \end{equation}
is satisfied for $\forall i,j \in \{1,2,\dots,\bar{\delta}\}$, then the buffered NCS is asymptotically stable for the control law given by \eqref{controller}. 

\textit{Proof:}
The LMIs \eqref{ineq2} can analogously to Theorem 1 be brought to the form $\bm{Q}_{ij}$ as in \eqref{Qij} with $\gamma_i$. Evaluating $\sum_{i = 1}^{\bar{\delta}} \alpha_{i,k}\sum_{j = 1}^{\bar{\delta}} \alpha_{j,k} \bm{Q}_{ij} $ from \eqref{Qij} leads to 
\begin{equation}
	\begin{bmatrix}
		\bm{\hat{P}} & \left( \bm{\hat{A}}(\bm{\alpha}_k) - \bm{\hat{B}} \bm{{\hat{K}}} \right)^T \bm{\hat{P}^+} \\
		\bm{\hat{P}^+} \left( \bm{\hat{A}}(\bm{\alpha}_k) - \bm{\hat{B}} \bm{{\hat{K}}} \right) & \left(1-\sum_{i = 1}^{\bar{\delta}}\alpha_{i,k}\gamma_i\right) \bm{\hat{P}^+}
	\end{bmatrix} > 0 
\end{equation}
from which the Schur complement can be computed
\begin{equation}
		\label{delgamma2}
\small \bm{\hat{P}} - \left( \bm{\hat{A}}(\bm{\alpha}_k) - \bm{\hat{B}} \bm{{\hat{K}}} \right)^T \bm{\hat{P}^+} \left( \bm{\hat{A}}(\bm{\alpha}_k)-\bm{\hat{B}} \bm{{\hat{K}}} \right)  > \sum_{i = 1}^{\bar{\delta}}\alpha_{i,k}\gamma_i \bm{\hat{P}}.
\end{equation}
 The equilibrium of \eqref{closed_lifted} is globally uniformly asymptotically stable if $\left| {\Delta}V(\bm{\xi}_k,\bm{\alpha}_k) \right| <0 $ which is, by the definition of the matrix $\bm{\hat{P}}$ and scalars $\gamma_i$,  ensured if \eqref{delgamma2} holds. \hfill $\blacksquare$
 
By varying parameters $\gamma_i$ ($i = 1,2,\dots,\bar{\delta}$) the lower bound for the transient decay rate of the system states can be increased possibly leading to even faster transient performance as in the case of one scalar $\gamma$ in Theorem 1.
 
\noindent Please note that control laws proposed in Theorems 2 and 3 can be extended analogously to the Theorem 4. Furthermore, the additional degrees of freedom $\gamma_i$ ($i = 1,2,\dots, \bar{\delta}$) represent only a slight modification of the implemented optimization problem since they don't introduce any new optimization variables or increase the number of LMIs.

\section{Illustrative Example}

The performance and efficiency of the proposed algorithms with respect to the reduced complexity are highlighted using a simulation example for a rotary servo plant described by a continuous-time plant. By choosing the angle $x_1 = \varphi(t)$ and rotational speed $x_2 = \dot{\varphi}(t) = \dot{x}_1$ as the state variables, the mathematical model \eqref{cont.plant} with  
	\begin{equation}
		\bm{A}_c = \begin{bmatrix*}
			0&1\\ 0&-\left( \Theta k_Gk_M+\dfrac{k_R}{J_M} \right)
		\end{bmatrix*}, \qquad 
		\bm{b}_c = 
		\renewcommand\arraystretch{1.5}
		\begin{bmatrix*}
			0\\ \Theta
		\end{bmatrix*}
	\label{exampleAB}
	\end{equation}
and $\Theta = \frac{\eta k_Gk_M}{J_MR_M}$ is derived \cite{Stanojevic2022ccta}. The values of the parameters for \eqref{exampleAB} are listed in Table \ref{param}. 
\begin{table}
	\medskip
	\caption{Parameters for the considered example \cite{Stanojevic2022ccta}}
	\setlength{\tabcolsep}{0.7\tabcolsep}
	\centering
	\begin{tabular}{ *{4}{l} }
		\toprule
		{Parameter} & {Symbol} & {Value} & Unit\\
		\midrule
		Armature resistance & $R_M$   &   $2.6 $&$\Omega$  \\
		Motor inertia & $J_M$ & $2.08 \times 10^{-3}$ &kgm$^2$ \\
		Viscous friction servo & $k_R$ & $48 \times 10^{-3}$ & Nm/s \\
		Motor back-emf constant & $k_M$           & $9.37 \times 10^{-3}$& V/s     \\
		Gear ratio &  $k_G$ &$70$&-\\
		Motor and gearbox efficiency &  $\eta$ &$0.621$&-\\
		\bottomrule
	\end{tabular}
	\label{param}
\end{table}
The sampling time of $T_d = 20$ms is chosen and the network induced delays are bounded by $0\leq {\tau}_k\leq 4T_d, \forall k$. Furthermore, the used network ensures the safe transmission of data, i.e. no dropouts occur, leading to $\bar{d} = 4$, $\bar{p} = 0$ and hence $\bar{\delta} = 4$. The considered NCS can therefore be described by a discrete-time model \eqref{disc.plant_buffer} with $
	\bm{A}_d  = \begin{bmatrix*}
		1   &  0.0106\\
		0  &  0.2347
	\end{bmatrix*}$ and $
	\bm{b}_d  = \begin{bmatrix*}
		0.0098\\
		0.7953
	\end{bmatrix*}$.
\subsection{Simulation study with one constant $\gamma$}
In the first simulation study, the existing approach from \cite{Cloosterman2010} is compared with the proposed control laws from Theorems 1 and 3 for one $\gamma$. Figure \ref{Fig1} presents the simulation results obtained for an initial state vector $\bm{x}_0 = \begin{bmatrix}
1&0
\end{bmatrix}^T$. 
\begin{figure}
	\medskip
	\centering
	\begin{tikzpicture}
		
		\definecolor{color1}{rgb}{0, 0.4470, 0.7410}
		\definecolor{color2}{rgb}{0.8500, 0.3250, 0.0980}
		\definecolor{color3}{rgb}{0.9290, 0.6940, 0.1250}
		\definecolor{color4}{rgb}{0.4940, 0.1840, 0.5560}
		\definecolor{color5}{rgb}{0.4660, 0.6740, 0.1880}
		\definecolor{color6}{rgb}{0.3010, 0.7450, 0.9330}
		\definecolor{color7}{rgb}{0.6350, 0.0780, 0.1840}
		
		
		\begin{axis}[
			xmin = 0,
			xmax = 2.5,
			height=4cm,
			width = 1.0\linewidth, 
			grid = both,
			y label style={font=\footnotesize,yshift=-0.45cm},
			ylabel={$x_{1,k}$},
			xticklabels = {},
			yticklabel style = {font=\footnotesize,xshift=0.5ex},
			]
			\addplot +[const plot, mark=none ,color1, solid, line width=1pt] table [x=t1, y=x11]{figures/sim1_x1.dat};
			\addplot +[const plot, mark=none ,color2, solid, line width=1pt] table [x=t2, y=x21]{figures/sim1_x2.dat};
			\addplot +[const plot, mark=none ,color4, densely dashdotted, line width=1pt] table [x=t3, y=x31]{figures/sim1_x3.dat};
		\end{axis}
		
		\begin{axis}[
			yshift = -2.5cm,
			xmin = 0,
			xmax = 2.5,
			width = 1.0\linewidth, 
			height  = 4cm,
			grid = both,
			y label style={font=\footnotesize,yshift=-0.5cm},
			x label style={font=\footnotesize},
			ylabel={$x_{2,k}$},
			xlabel={$t$ in s},	
			yticklabel style = {font=\footnotesize,xshift=0.5ex},
			xticklabel style = {font=\footnotesize},
			legend columns=1, 
			legend entries={ \cite{Cloosterman2010}: \,\,\,\,\,$u_k = -\bm{K}_x \bm{x}_k$ $\gamma = 0.075$,Th 1: $u_k = -\bm{K}_x \bm{x}_k$ $\gamma = 0.089$,Th 4: $u_k = - [\bm{K}_x \,\,\,\, \bm{K}_u]\bm{\xi}_k$ $\gamma = 0.31$},
			legend style={font=\footnotesize,at={(0.995,0.01)},anchor=south east},
			legend cell align={left}
			]
			\addplot +[const plot, mark=none ,color1, solid, line width=1pt] table [x=t1, y=x12]{figures/sim1_x1.dat};
			\addplot +[const plot, mark=none ,color2, solid, line width=1pt] table [x=t2, y=x22]{figures/sim1_x2.dat};
			\addplot +[const plot, mark=none ,color4, densely dashdotted, line width=1pt] table [x=t3, y=x32]{figures/sim1_x3.dat};
		\end{axis}	
	\end{tikzpicture}	
	\caption{Simulation results for $\bm{x}_k$ with the control laws from Table \ref{sim1} }
	\label{Fig1}
\end{figure}
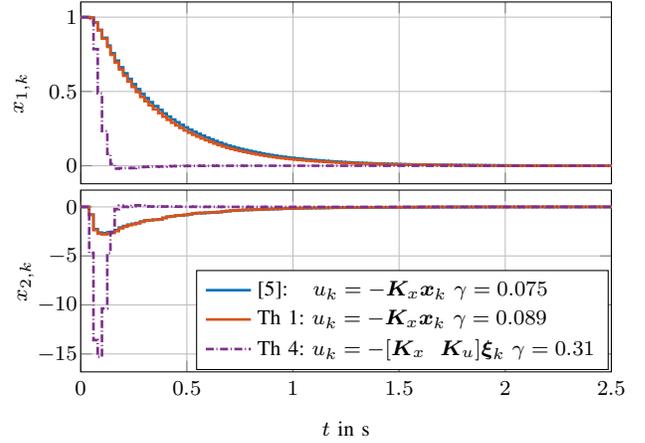
Table \ref{sim1} shows an overview of the obtained control laws for the largest admissible value of $\gamma$ and therefore the fastest transient decay rate for which a feasible solution can be obtained. Additionally, in order to compare the computational effort, the number of optimization variables, the number of LMIs of the form \eqref{LMIimplement} and the time needed to solve the optimization problem (Intel(R) Core(TM) i7-8565U CPU $@$ 1.80GHz) are listed as well. 

\begin{table}
	\caption{Figure \ref{sim1} - Computational evaluation for one $\gamma$}
	\setlength{\tabcolsep}{1\tabcolsep}
	\centering
	\begin{tabular}{ |l  l|| c | c|c | }
		\hline
		 \multicolumn{2}{|c||}{\multirow{2}{*}{Control Law}} &  \multirow{2}{*}{$\gamma$} &\makecell{opt.var.}  &  \multirow{2}{*}{{Time\footnotetext[1]{Footnote}}} \\ \cline{4-4}
		   						  & &  &LMIs  & in $s$\\ \cline{4-5} \hline

		\renewcommand{\arraystretch}{18} \cite{Cloosterman2010} &  $\begin{aligned}
			\bm{u}_k &= - \begin{bmatrix}
				\bm{K}_x & \bm{0}
			\end{bmatrix} \xi_k  \\
			\bm{K}_x &=  \begin{bmatrix}
				2.62 &   0.04
			\end{bmatrix}  
		\end{aligned}$ & 0.075 & \makecell{11526 \\ 65536} & 4648.7 \\ \hline
		Th 1 &$\begin{aligned}
			\bm{u}_k &= - \begin{bmatrix}
				\bm{K}_x & \bm{0}
			\end{bmatrix} \xi_k  \\
			\bm{K}_x &=  \begin{bmatrix}
				2.73  &  0.034
			\end{bmatrix}  
		\end{aligned}$   &0.089  & \makecell{186 \\ 16} & 0.39 \\ \hline 
		Th 3 &$\begin{aligned}
			\bm{u}_k &= - \begin{bmatrix}
				\bm{K}_x & \bm{K}_u
			\end{bmatrix} \xi_k  \\
			\bm{K}_x &=  \begin{bmatrix}
				15.69  &   0.20
			\end{bmatrix}  \\
		\bm{K}_u &=  \begin{bmatrix}
		0.36 \,\,\,   0.22 \,\,\,   0.17 \,\,\,  0.13
	\end{bmatrix}  
		\end{aligned}$ &0.31  &\makecell{118 \\ 16} & 0.27\\ \hline
	\end{tabular}
	\label{sim1}
\end{table}

The control law proposed in Theorem 1 shows slightly faster transient decay rate of the system states in comparison to the existing approach from \cite{Cloosterman2010}. The extended formulation of the control algorithm (purple curve) can yield much faster response as with the static control methods, but, as previously mentioned, additional restrictions must hold. The most striking aspect of the comparison can be seen in Table \ref{sim1} in the number of optimization variables and LMIs as well as the time needed to solve the corresponding optimization problem. Not only the complexity with respect to necessary over-approximation techniques from \cite{Cloosterman2010} is avoided by simple reformulation of the buffered NCS as a switched system, but the resulting optimization problem is greatly simplified. The computation effort is therefore significantly reduced and the solution can be obtained in the matter of seconds.

\noindent Please note, that increasing the value of $\bar{\delta}$ to 5 would result in 46086 optimization variables and 1048576 LMIs necessary to obtain the control law proposed in \cite{Cloosterman2010}, which greatly increases the complexity for the over-approximation of the NCS and computational effort for the control synthesis. For the proposed control law from e.g. Theorem 1, the value $\bar{\delta} = 5$ would, however, lead to only 321 optimization variables and 25 LMIs. The complexity of defining the NCS as switched system and the corresponding computational effort in this case would not be affected at all.
\subsection{Simulation study with multiple $\gamma_i$}
Furthermore, to underpin the impact of additional degrees of freedom in the control design, the control laws from Theorems 1 and 2 with $\bm{\gamma} = \begin{bmatrix}
\gamma_1&\gamma_2&\gamma_3&\gamma_4
\end{bmatrix}^T$ obtained for an initial state vector $\bm{x}_0 = \begin{bmatrix} 
	1&0
\end{bmatrix}^T$ 
are compared with the existing approach from \cite{Cloosterman2010}. Figure \ref{Fig2} and Table \ref{sim2} summarize the results for the compared approaches. 

\begin{figure}
	\centering
	\medskip
		\begin{tikzpicture}
		
		\definecolor{color1}{rgb}{0, 0.4470, 0.7410}
		\definecolor{color2}{rgb}{0.8500, 0.3250, 0.0980}
		\definecolor{color3}{rgb}{0.9290, 0.6940, 0.1250}
		\definecolor{color4}{rgb}{0.4940, 0.1840, 0.5560}
		\definecolor{color5}{rgb}{0.4660, 0.6740, 0.1880}
		\definecolor{color6}{rgb}{0.3010, 0.7450, 0.9330}
		\definecolor{color7}{rgb}{0.6350, 0.0780, 0.1840}
		
		
		\begin{axis}[
			xmin = 0,
			xmax = 2.5,
			height=4cm,
			width = 1.0\linewidth, 
			grid = both,
			y label style={font=\footnotesize,yshift=-0.45cm},
			ylabel={$x_{1,k}$},
			xticklabels = {},
			yticklabel style = {font={ \footnotesize},xshift=0.5ex},
			]
			\addplot +[const plot, mark=none ,color1, solid, line width=1pt] table [x=t1, y=x11]{figures/sim1_x1.dat};
			\addplot +[const plot, mark=none ,color2, solid, line width=1pt] table [x=t2, y=x21]{figures/sim2_x2.dat};
			\addplot +[const plot, mark=none ,color4, densely dashdotted, line width=1pt] table [x=t3, y=x31]{figures/sim2_x3.dat};
		\end{axis}
		
		\begin{axis}[
			yshift = -2.5cm,
			xmin = 0,
			xmax = 2.5,
			ymin = -7,
			width = 1.0\linewidth, 
			height  = 4cm,
			grid = both,
			y label style={font=\footnotesize,yshift=-0.5cm},
			x label style={font=\footnotesize},
			ylabel={$x_{2,k}$},
			xlabel={$t$ in s},	
			yticklabel style = {font=\footnotesize,xshift=0.5ex},
			xticklabel style = {font=\footnotesize},
			legend style={font=\footnotesize,at={(0.995,0.01)},anchor=south east},
			legend columns=1, 
			legend entries={\cite{Cloosterman2010}: $u_k = -\bm{K}_x \bm{x}_k$ $\gamma = 0.075$,Th 4: $u_k = -\bm{K}_x \bm{x}_k$ $\gamma = [0.19 \,\, {0} \,\, 0 \,\, 0]$,
				Th 2$\&$4: $u_k = -\bm{K}_{x,i} \bm{x}_k$ $\gamma = [0.225 \,\, {0} \,\, 0 \,\, 0]$},
			legend style={row sep=-0.2pt},
			legend cell align={left}
			]
			\addplot +[const plot, mark=none ,color1, solid, line width=1pt] table [x=t1, y=x12]{figures/sim1_x1.dat};
			\addplot +[const plot, mark=none ,color2, solid, line width=1pt] table [x=t2, y=x22]{figures/sim2_x2.dat};
			\addplot +[const plot, mark=none ,color4, densely dashdotted, line width=1pt] table [x=t3, y=x32]{figures/sim2_x3.dat};
		\end{axis}	
	\end{tikzpicture}
	\caption{Simulation results for $\bm{x}_k$ with the control laws from Table \ref{sim2} }
	\label{Fig2}
\end{figure}
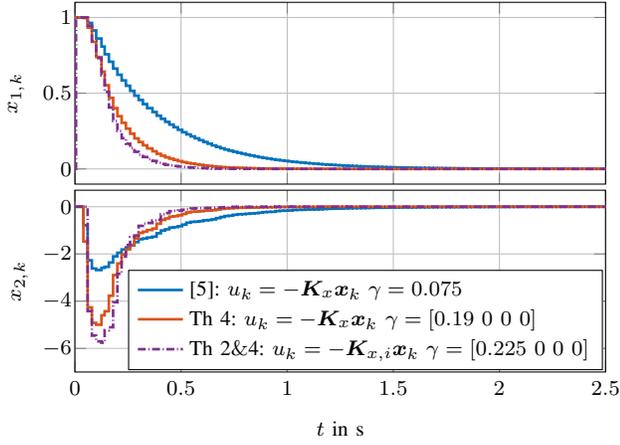
\begin{table}
	\caption{Figure \ref{sim2} - Computational evaluation for multiple $\gamma_i$}
	\setlength{\tabcolsep}{1\tabcolsep}
	\begin{tabular}{ |l  l|| c | c|c | }
	\hline
	\multicolumn{2}{|c||}{\multirow{2}{*}{Control Law}} &  \multirow{2}{*}{$\gamma$} &\makecell{opt.var.}  &  \multirow{2}{*}{{Time\footnotetext[1]{Footnote}}} \\ \cline{4-4}
	& &  &LMIs  & in s\\ \cline{4-5} \hline
		\renewcommand{\arraystretch}{18} \cite{Cloosterman2010} &  $\begin{aligned}
		\bm{u}_k &= - \begin{bmatrix}
			\bm{K}_x & \bm{0}
		\end{bmatrix} \xi_k  \\
		\bm{K}_x &=  \begin{bmatrix}
			2.62 &   0.04
		\end{bmatrix}  
	\end{aligned}$ & 0.075 &\makecell{11526 \\ 65536} & 4648.7\\ \hline
		Th 4 &$\begin{aligned}
			\bm{u}_k &= - \begin{bmatrix}
				\bm{K}_x & \bm{0}
			\end{bmatrix} \xi_k  \\
			\bm{K}_x &=  \begin{bmatrix}
				4.93  &  0.09
			\end{bmatrix}  
		\end{aligned}$   & $\begin{bmatrix}
		0.19 \\0 \\0\\0
	\end{bmatrix}$  &\makecell{186 \\ 16} & 0.32\\ \hline 
		$\begin{aligned}
			&\text{Th 4} \\ &{\footnotesize \text{switched}}\\ &{\footnotesize\text{control }}\\ &{\footnotesize\text{law}}
		\end{aligned}$ &$\begin{aligned}
			&\bm{u}_k = - \begin{bmatrix}
				\bm{K}_{x,i} & \bm{0}
			\end{bmatrix} \xi_k  \\
			&\bm{K}_{x,1} =  \begin{bmatrix}
				5.56  &  0.086
			\end{bmatrix}  \\
		&\bm{K}_{x,2} =  \begin{bmatrix}
		5.6  &  0.08
	\end{bmatrix} \\
	&\bm{K}_{x,3} =  \begin{bmatrix}
	5.87  &  0.08
	\end{bmatrix} \\
	 &\bm{K}_{x,4}=  \begin{bmatrix}
		6.09  &  0.096
	\end{bmatrix}
		\end{aligned}$ & $\begin{bmatrix}
		0.225 \\0 \\0\\0
	\end{bmatrix}$  & \makecell{204 \\ 16} & 0.24\\ \hline
	\end{tabular}
	\label{sim2}
\end{table}
\noindent Figure \ref{Fig2}  illustrates the great effect different $\gamma_i$  ($\forall i = 1,2,3,4$) can have on the transient decay rate of the system states. Already in the case of a static controller, the specific choice of $\gamma_i$ leads to the further increase of the system states decay rate. The even more significant improvement in the convergence speed  is achieved using a switching control law, where the buffer applies the control signal depending on the active switching mode. The computation effort and time needed remain the same as in the case with a single $\gamma$.

\section{Conclusion and Outlook}
In this paper, a control strategy for NCS affected by random variable delays and packet dropouts with a special focus on keeping the computational complexity of the control design as low as possible is proposed. This is achieved by first introducing a new buffering mechanism, which on the one hand simplifies the discrete-time model of the NCS significantly by ensuring that there is only one control signal acting between two sampling instants. On the other hand, it limits the additional delay introduced by the buffer to one sampling time. The resulting buffered NCS can be therefore easily described as a switched system, which reduces the complexity of the control design greatly, since procedures  such as Jordan Form for the over-approximation of the NCS model to obtain a polytopic model suitable for stability analysis as in \cite{PosthumusCloosterman2008}, \cite{Cloosterman2010} are no longer necessary. Hence, the stability conditions and controller synthesis can be defined in terms of strongly reduced number of LMIs. The performance and computational effort of the proposed control approaches is evaluated based on a simulation example which not only demonstrates greater flexibility of the proposed laws with respect to the transient behavior but highlights the achieved computational simplicity as well. 

\noindent Future developments will focus on the impact on the transient behavior and systematic choice for $\gamma_i$. Furthermore, a generalization of the proposed strategies to more complex multi-hop or\,\,spatially distributed network structures is of interest. 
\bibliographystyle{IEEEtran}
\bibliography{Katarina}

\end{document}